\documentclass[aps,prb,groupedaddress, twocolumn]{revtex4-1}
\usepackage{amsmath}
\usepackage{amssymb}
\usepackage{braket}
\usepackage{mathtools} 
\usepackage[pdftex]{graphicx}
\usepackage[space]{grffile} 
\graphicspath{{figs/}{figs/diagrams/}}

\usepackage[usenames]{color} 

\definecolor{jade}{rgb}{0.0, 0.66, 0.42}
\definecolor{jasper}{rgb}{0.84, 0.23, 0.24}
\definecolor{lapislazuli}{rgb}{0.15, 0.38, 0.61}

\usepackage{mathptmx} 
\usepackage[unicode=true,pdfusetitle,
 bookmarks=true,bookmarksnumbered=false,bookmarksopen=false,
 breaklinks=false,pdfborder={0 0 1},backref=false,colorlinks=true,citecolor=lapislazuli,
 urlcolor=lapislazuli,linkcolor=lapislazuli]{hyperref}

\AtBeginDocument{\usepackage{booktabs}}
\usepackage{array}
\newcolumntype{C}{>{$\displaystyle}c<{$}}

\newcommand{\abs}[1]{\left| #1  \right| }
\newcommand{\avg}[1]{\left< #1  \right> }
\newcommand{\free}[1]{\left\{ #1  \right\}} 

\newcommand{\Tr}[1]{\text{Tr}\left( #1 \right)}

\makeatletter
\def\l@subsubsection#1#2{}
\makeatother

\begin{document}

\title{Hamiltonian dynamics of a sum of interacting random matrices}

\author{Matteo Bellitti}
\author{Siddhardh Morampudi}
\author{Chris R. Laumann}

\affiliation{Department of Physics, Boston University, Boston, MA 02215, USA}


\begin{abstract}
In ergodic quantum systems, physical observables have a non-relaxing component if they ``overlap'' with a conserved quantity. 
In interacting microscopic models, how to isolate the non-relaxing component is unclear.
We compute exact dynamical correlators governed by a Hamiltonian composed of two large interacting random matrices, $H=A+B$. 
We analytically obtain the late-time value of $\langle A(t) A(0) \rangle$; this quantifies the non-relaxing part of the observable $A$. 
The relaxation to this value is governed by a power-law determined by the spectrum of the Hamiltonian $H$, independent of the observable $A$. 
For Gaussian matrices, we further compute out-of-time-ordered-correlators (OTOCs) and find that the existence of a non-relaxing part of $A$ leads to modifications of the late time values and exponents.
Our results follow from exact resummation of a diagrammatic expansion and hyperoperator techniques.
\end{abstract}

\maketitle

\tableofcontents

\section{Introduction}

Consider an isolated quantum system whose Hamiltonian can be decomposed into two terms
\begin{align}
\label{eq:hamab}
    H = A + B
\end{align}
Under what conditions does the observable $A$ have a non-relaxing component? 
There are many more or less exotic mechanisms for $\langle A(t) A(0) \rangle_c$ to approach a non-zero constant at late time: $A$ could commute with $H$, or the whole system could be localized\cite{Anderson1958, GornyiLocalization2005, BAA2006, Serbynlbit2013, Huselbit2014, NandkishoreReview2015}, scarred\cite{BernienScars2017, TurnerScars2018, MotrunichScars2019, ChoiScars2019}, shattered\cite{KhemaniShattered2019}, or otherwise mistreated by theorists.
However, even in the simple case in which $H$ is ergodic and $A$ does not commute with it, the operator $A$ appears to be ``part'' of the conserved energy and we expect it to have a non-relaxing component. 

This intuition is misleading. 
For any operator $O$ and any Hamiltonian $H$, one could write $H = (H-O) + O$ and formally reproduce the decomposition in Eq.~\eqref{eq:hamab} without learning anything.
For Eq.~\eqref{eq:hamab} to be nontrivial, there must be some physical sense in which the decomposition is defined.
In this paper, we consider the case in which $A$ and $B$ are large independent random matrices of dimension $N$ with arbitrary spectra. 
They need not obey Wigner's celebrated semicircle law, but the eigenvectors of $A$ must be in generic position with respect to those of $B$. 
While this choice appears technical, it is a natural model if $H$, $A$ and $B$ each satisfy the eigenstate thermalization hypothesis (ETH) and $A$ and $B$ are built out of physically distinct local operators.

Our primary result is an exact integral representation of the dynamical correlator $\langle A(t) A(0)\rangle_c$ at large~$N$ (Eq.~\eqref{eq:ng_freq_corr}). 
From this representation we derive the asymptotic structure at late time,
\begin{equation}
    \label{eq:long_time_form_g}
  \avg{A(t) A(0)}_c \underset{t\to\infty}{\sim} A_\infty^2  + C_{A,H} \abs{\avg{e^{i H t}}}^2 
\end{equation}
This form has two important features:
first, the late-time constant $A_\infty^2$ quantifies the non-relaxing component of $A$. 
It has a definite integral representation in terms of the densities of states of $A$ and $H$ (see Eq.~\eqref{eq:constant}). 
In general, $A_\infty^2$ is not zero --this accords with the intuition that $A$ is `part' of the energy-- however it is not given by a simple trace overlap.

Second, the $\langle A(t) A(0) \rangle_c$ approaches its late-time value with a form governed by the characteristic function of $H$ alone. 
The particular operator $A$ only enters through the constant rescaling $C_{A,H}$. 
This implies that the late time power laws come from the singularities in the density of states of $H$ and do not depend on any detailed structure of $A$. 
For the many random matrix ensembles where the density of states of $H$ has square-root singularities --the semicircle law is one example-- Eq.~\eqref{eq:long_time_form_g} predicts $1/t^3$ decay. 
Still, other possibilities exist: if $A$ and $B$ are projectors, for example, the resulting Hamiltonian has different singularities which produce slower $1/t$ decay.

We set up the calculation at infinite temperature, but once we have those results the full temperature dependence can be derived introducing Boltzmann weights in appropriate places (Sec.\ref{sec:finite_temperature}).
In the zero temperature limit the power law approach of Eq.~\eqref{eq:long_time_form_g} is modified and typically the decay exponents are reduced by a factor of~$2$. This behavior is reminiscent of the SYK model\cite{KitaevTalk2015, Kitaev2018, MaldacenaSYK2016, cotlerBlackHolesRandom2017, AltlandSYK2016} at times $t>N$, the number of fermions, but $t < e^{Ns}$, the level spacing.

In the special case where $A$ and $B$ are both Gaussian random matrices, and thus satisfy the semi-circle law, these results simplify significantly:
the late time value $A_\infty^2$ is given by a trace overlap between $A$ and $H$ (Eq.~\eqref{eq:gauss_asymp_form}), as one might have guessed, and the asymptotic form Eq.~\eqref{eq:long_time_form_g} holds exactly for all times $t$. 
We derive these simplified forms both from our general diagrammatic formalism, and from a more direct hyperoperator approach.

There has been a lot of recent interest in computing out of time ordered correlators (OTOCs) as a rough characterization of quantum chaos\cite{KitaevTalk2015, Kitaev2018, Nahum2018, Maldacena2016, KhemaniConserved2018, Keyserlingk2018, vijayFiniteTemperatureScramblingRandom2018, ChalkerChaos2018}. 
For Gaussian $A$ and $B$ we use the hyperoperator technique to compute the OTOC,
\begin{align}
    \frac{1}{2} \langle [A(t), A(0)]^2 \rangle
\end{align}
and find that it decays asymptotically more slowly than if it did not contain a conserved piece.

It is worthwhile noting that much recent analytical progress on dynamics in many-body systems has been made in the setting of random unitary circuits.\cite{ChandranClifford2015, Nahum2017, Nahum2018, Keyserlingk2018} These models do not naturally have conservation laws, and introducing the extra structure to create them\cite{KhemaniConserved2018, RakovszkyConserved2018} makes calculations significantly harder. On the contrary, Hamiltonian models automatically come with a conserved energy, but deriving exact results there is difficult in the absence of further structure like conformal symmetry, integrability, or a large-$N$ limit.
Our work falls in this last class: we treat a Hamiltonian system exactly, at the price of introducing large-$N$ random matrices.

All of our results are at infinite temperature and averaged over the random matrix ensemble; for the rest of the paper, the ``expectation value'' symbol means
\begin{equation}
    \label{eq:free_prob_functional}
    \avg{ \circ } \coloneqq \frac{1}{N} \mathbb{E}_{A,B} \left[  \Tr{ \circ } \right]
\end{equation}
The normalization is such that the $N \times N$ identity matrix has expectation equal to $1$.

The ensemble for $A$ and $B$ is fairly general: we just need their densities of states $\rho_A$ and $\rho_B$ to be well defined in the limit $N \to \infty$, and their eigenspaces to be in generic position. One way to realize such matrices is
\begin{equation}
    A = U^\dagger \Lambda U 
\end{equation}
where $\Lambda$ is a diagonal matrix and $U$ is a Haar unitary.

The paper is organized as follows: 
in Sec.\ref{sec:exact_analysis} we introduce a set of diagrammatic tools and use them 
to derive the exact frequency space two--point function.
Using this result we study (Sec.\ref{sec:asymptotics}) the long time behavior of
$\langle A(t)A(0) \rangle$ and identify the structure mentioned in Eq.\eqref{eq:long_time_form_g}.
We then discuss (Sec.\ref{sec:finite_temperature}) how these results are modified at finite temperature. 
In Sec.\ref{sec:orthogonal_transformation} we introduce the direct hyperoperator approach for Gaussian matrices. We use it to reproduce the results of the previous sections and compute the OTOC.
Finally, in Sec.\ref{sec:numerical_confirmation} we compare the predictions of our analysis with numerics, obtaining a satisfactory agreement.

\section{Exact Analysis} 
\label{sec:exact_analysis}

\subsection{Frequency space representation of correlators}
\label{sec:frequency_space_representation}

\begin{figure}
\centering
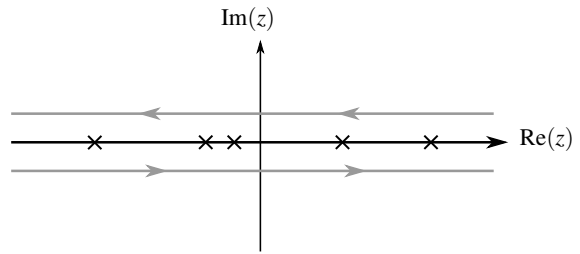
\caption{\label{fig:cauchy_contour} Integration contour for the Cauchy transform in eq.~\ref{eq:cauchy_transform}. The crosses are eigenvalues of $H$.}
\end{figure}

The correlation function we focus on in this section is
\begin{equation}
  \label{eq:2_correlator}
  G(t) = \avg{A(t)A(0)} = \avg{e^{iHt} A e^{-iHt} A}
\end{equation}
It is convenient to reformulate this problem in frequency space, using the Cauchy representation of the time evolution operator:
\begin{equation}
  \label{eq:cauchy_transform}
  e^{iHt} = \oint \frac{dz}{2 \pi i} \frac{e^{izt}}{z - H}
\end{equation}
where the integral is over any contour that encloses the full spectrum of $H$ (see fig.~\ref{fig:cauchy_contour}). At any finite Hilbert space size $N$ this contour is closed, and in the thermodynamic limit it can be closed at infinity.

Using the integral representation just introduced, we can rewrite eq.~\ref{eq:2_correlator} as
\begin{align}
  G(t) &= \oint \frac{dz}{2 \pi i} \frac{dw}{2 \pi i} e^{i(z-w)t} \avg{\frac{1}{z-H} A \frac{1}{w-H} A} \\
\label{eq:inverse_cauchy} &= \oint \frac{dz}{2 \pi i} \frac{dw}{2 \pi i} e^{i(z-w)t} G(z,w)
\end{align}
which defines the frequency space correlator:
\begin{align}
  \label{eq:resolvent_series}
  G(z,w) &\coloneqq \avg{\frac{1}{z-H} A \frac{1}{w-H} A} \\
  &= \sum_{m=0}^\infty \sum_{n=0}^\infty \frac{1}{z^{n+1}} \frac{1}{w^{m+1}}\avg{H^n A H^m A}
\end{align}
The last formal manipulation involving the geometric series shows that we can reduce the problem of determining $G(z,w)$ to the calculation of mixed moments of $A$ and $H$:
\begin{equation}
  \label{eq:freq_correlator}
  \avg{H^n A H^m A} = \avg{ \left(A + B\right)^n A  \left(A + B\right)^m A}
\end{equation}
For practical purposes it is better to write the calculation in terms of the
matrices $A$ and $B$, as they are the independent objects in the problem.

If $A$ and $B$ are Gaussian random matrices, the evaluation of the moments is readily accomplished using Wick's theorem. 
For the more general (non-semicircle) matrices we would like to treat here, the \emph{free cumulant expansion} plays the role of a generalized Wick's theorem for non-commuting matrices.
Given a set of noncommuting random variables $X_i$, we can recursively define their \emph{free cumulants}
(denoted with curly braces) through the formula
\begin{equation}
\label{eq:free_cumulant_expansion}
\avg{X_1 \ldots X_n} \eqqcolon \sum_{\pi \in \text{NC}(n)} \prod_{b \in \pi}
\left\{ X_{i_1} \ldots X_{i_{|b|}} \right\}
\end{equation}
where $\text{NC(k)}$ denotes the set non--crossing partitions of $k$ objects,
and $b$ is a block in the partition $\pi$. 
See Ref.~\onlinecite{morampudiManybodySystemsRandom2018} for more pedagogical details of this formalism. 
For aficionados of planar perturbation theory, we note that the free cumulants of a random matrix $X$ are closely related to the fully renormalized vertices of $X$. 

Restating the problem in terms of free cumulants makes things simpler: in the large $N$ limit all the mixed free cumulants of independent random
matrices vanish\cite{nicaLecturesCombinatoricsFree2006}, which simplifies enormously the calculation of moments like the one of eq.\ref{eq:freq_correlator}.  When two random matrices have this property, we say that the two variables are \emph{freely independent}.

\subsection{Diagrammatic Solution}
\label{sub:diagrammatic_rules}
The calculation of moments of large independent random matrices lends itself to
a convenient diagrammatic representation through the free cumulant expansion
(eq.~\ref{eq:free_cumulant_expansion}). 
The diagrammatics are similar to those of Refs.~\onlinecite{brezinPlanarDiagrams1993,brezinCorrelationFunctionsDisordered1994} although our interpretation in terms of free cumulants is somewhat more recent \cite{nicaLecturesCombinatoricsFree2006, morampudiManybodySystemsRandom2018}.

We are computing trace moments, so it is natural to represent them with a circular diagram.
Factors of $A$ will be represented by the insertion of a full dot $(\bullet)$
along the circle and $B$ factors by insertions of an empty one ($\circ$).

Notice that two of the $A$ insertions are not like the others: they do not come from the
expansion of $H^n$. Each diagram is then naturally split in two halves by those
special $A$ insertions, one half associated to the $H^n/z^{n+1}$ term and
the other one to the $H^m/w^{m+1}$ term.

Associate single lines (---) to factors of $1/z$ and $1/w$: lines in the upper semicircle
represent factors of $1/z$ and lines in the lower semicircle factors of $1/w$.

A vertex with double line legs (=) that connects a set of insertions
represents the free cumulant of those operators. Notice that all the insertions participating in a given vertex must be of the same kind, as mixed free cumulants of $A$ and $B$ vanish.

Let us summarize the diagrammatic rules we have just introduced:
\begin{center}
\begingroup%
  \makeatletter%
  \providecommand\color[2][]{%
    \errmessage{(Inkscape) Color is used for the text in Inkscape, but the package 'color.sty' is not loaded}%
    \renewcommand\color[2][]{}%
  }%
  \providecommand\transparent[1]{%
    \errmessage{(Inkscape) Transparency is used (non-zero) for the text in Inkscape, but the package 'transparent.sty' is not loaded}%
    \renewcommand\transparent[1]{}%
  }%
  \providecommand\rotatebox[2]{#2}%
  \newcommand*\fsize{\dimexpr\f@size pt\relax}%
  \newcommand*\lineheight[1]{\fontsize{\fsize}{#1\fsize}\selectfont}%
  \ifx\svgwidth\undefined%
    \setlength{\unitlength}{176.77431242bp}%
    \ifx\svgscale\undefined%
      \relax%
    \else%
      \setlength{\unitlength}{\unitlength * \real{\svgscale}}%
    \fi%
  \else%
    \setlength{\unitlength}{\svgwidth}%
  \fi%
  \global\let\svgwidth\undefined%
  \global\let\svgscale\undefined%
  \makeatother%
  \begin{picture}(1,0.13014573)%
    \lineheight{1}%
    \setlength\tabcolsep{0pt}%
    \put(0,0){\includegraphics[width=\unitlength,page=1]{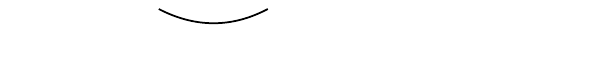}}%
    \put(0.31198264,0.00401619){\color[rgb]{0,0,0}\makebox(0,0)[lt]{\lineheight{1.25}\smash{\begin{tabular}[t]{l}$1/w$\end{tabular}}}}%
    \put(0,0){\includegraphics[width=\unitlength,page=2]{rules.pdf}}%
    \put(0.05779994,0.00401619){\color[rgb]{0,0,0}\makebox(0,0)[lt]{\lineheight{1.25}\smash{\begin{tabular}[t]{l}$1/z$\end{tabular}}}}%
    \put(0,0){\includegraphics[width=\unitlength,page=3]{rules.pdf}}%
    \put(0.51414852,0.00470668){\color[rgb]{0,0,0}\makebox(0,0)[lt]{\lineheight{1.25}\smash{\begin{tabular}[t]{l}$A$\end{tabular}}}}%
    \put(0,0){\includegraphics[width=\unitlength,page=4]{rules.pdf}}%
    \put(0.63378741,0.00470668){\color[rgb]{0,0,0}\makebox(0,0)[lt]{\lineheight{1.25}\smash{\begin{tabular}[t]{l}$B$\end{tabular}}}}%
    \put(0.77923159,0.00470674){\color[rgb]{0,0,0}\makebox(0,0)[lt]{\lineheight{1.25}\smash{\begin{tabular}[t]{l}$\left\{ \ldots \right\}$\end{tabular}}}}%
    \put(0,0){\includegraphics[width=\unitlength,page=5]{rules.pdf}}%
  \end{picture}%
\endgroup%

\end{center}

Finally, in the free cumulant expansion we only need noncrossing partitions, so these diagrams are \emph{planar}.
The planar nature of these diagrams makes them easy to classify,
which will be very useful in sections
\ref{sub:propagator_renormalization} and \ref{sub:box_diagram}
where we compute a perturbative resummation of infinitely many
diagrams.

Our aim is the following: to compute $G(z,w)$ we must \emph{sum
over all planar diagrams with at least two $A$ insertions and any
number of $A$ and $B$ insertions in the upper and lower halves}.
There are no extra combinatorial factors.

As an example of the application of these rules, let us calculate a diagram that will be included in the expansion of $G(z,w)$:
\begin{equation}
\begingroup%
  \makeatletter%
  \providecommand\color[2][]{%
    \errmessage{(Inkscape) Color is used for the text in Inkscape, but the package 'color.sty' is not loaded}%
    \renewcommand\color[2][]{}%
  }%
  \providecommand\transparent[1]{%
    \errmessage{(Inkscape) Transparency is used (non-zero) for the text in Inkscape, but the package 'transparent.sty' is not loaded}%
    \renewcommand\transparent[1]{}%
  }%
  \providecommand\rotatebox[2]{#2}%
  \newcommand*\fsize{\dimexpr\f@size pt\relax}%
  \newcommand*\lineheight[1]{\fontsize{\fsize}{#1\fsize}\selectfont}%
  \ifx\svgwidth\undefined%
    \setlength{\unitlength}{124.53116469bp}%
    \ifx\svgscale\undefined%
      \relax%
    \else%
      \setlength{\unitlength}{\unitlength * \real{\svgscale}}%
    \fi%
  \else%
    \setlength{\unitlength}{\svgwidth}%
  \fi%
  \global\let\svgwidth\undefined%
  \global\let\svgscale\undefined%
  \makeatother%
  \begin{picture}(1,0.33068433)%
    \lineheight{1}%
    \setlength\tabcolsep{0pt}%
    \put(0,0){\includegraphics[width=\unitlength,page=1]{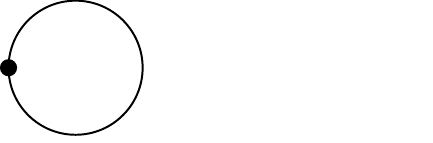}}%
    \put(-1.22594671,0.43002449){\color[rgb]{0,0,0}\makebox(0,0)[lt]{\begin{minipage}{1.04534932\unitlength}\raggedright \end{minipage}}}%
    \put(0.40694568,0.15900851){\color[rgb]{0,0,0}\makebox(0,0)[lt]{\lineheight{1.25}\smash{\begin{tabular}[t]{l}= $\dfrac{1}{z^3w^2}\free{A^2}\free{B^2}\free{B} $\end{tabular}}}}%
    \put(0,0){\includegraphics[width=\unitlength,page=2]{nontrivial_example.pdf}}%
    \put(0.43963026,0.41545303){\color[rgb]{0,0,0}\makebox(0,0)[lt]{\begin{minipage}{0.66033381\unitlength}\raggedright \end{minipage}}}%
  \end{picture}%
\endgroup%

\end{equation}
In this simple case there are only vertices with one or two legs,
but in general a vertex can have any number of legs.

Without loss of generality, for the rest of the paper we will assume that $\free{A} = \free{B} = 0$. This does not change the dynamics, as it corresponds to a constant energy shift, but makes the diagrammatics much simpler.

\subsubsection{Propagator}
\label{sub:propagator_renormalization}

The \emph{full propagator} of $H$, $f_H(z)$, is
\begin{equation}
  \label{eq:propagator_series}
  f_H(z) \coloneqq \avg{\frac{1}{z-H}} = \sum_{n=0}^\infty \frac{\avg{H^n}}{z^{n+1}}.
\end{equation}
Below, we suppress the subscript $H$ when there is no risk of confusion.
As usual, the complex analytic features of the propagator encode the spectrum of $H$. 
For example, the mean spectral density $\rho(x)$ can be obtained by inserting the Sokhotski--Plemelj formula
\begin{equation}
  \label{eq:sp}
  \lim_{ \delta \to 0^+} \frac{1}{x \pm i \delta} = P \frac{1}{x} \mp i \pi \delta(x)
\end{equation}
into the definition of the propagator,
\begin{align}
  f(x \pm i \delta) &= \avg{\frac{1}{x - H \pm i \delta}} \\
  &= \label{eq:prop_real_axis} \phi(x) \mp i \pi \rho(x)
\end{align}
Here, $\phi(x)$ is the Hilbert transform of $ \rho(x)$:
\begin{equation}
  \phi(x) \coloneqq \avg{P \frac{1}{x - H}} = P \int d\lambda \ \frac{\rho(\lambda)}{x - \lambda}
\end{equation} 

\subsubsection{1PI Diagrams} 
\label{sub:1pi_diagrams}

We write a diagrammatic representation of eq.~\ref{eq:propagator_series} by introducing a thick line to represent $f(z)$:
\begin{equation}
\begingroup%
  \makeatletter%
  \providecommand\color[2][]{%
    \errmessage{(Inkscape) Color is used for the text in Inkscape, but the package 'color.sty' is not loaded}%
    \renewcommand\color[2][]{}%
  }%
  \providecommand\transparent[1]{%
    \errmessage{(Inkscape) Transparency is used (non-zero) for the text in Inkscape, but the package 'transparent.sty' is not loaded}%
    \renewcommand\transparent[1]{}%
  }%
  \providecommand\rotatebox[2]{#2}%
  \newcommand*\fsize{\dimexpr\f@size pt\relax}%
  \newcommand*\lineheight[1]{\fontsize{\fsize}{#1\fsize}\selectfont}%
  \ifx\svgwidth\undefined%
    \setlength{\unitlength}{213.21731447bp}%
    \ifx\svgscale\undefined%
      \relax%
    \else%
      \setlength{\unitlength}{\unitlength * \real{\svgscale}}%
    \fi%
  \else%
    \setlength{\unitlength}{\svgwidth}%
  \fi%
  \global\let\svgwidth\undefined%
  \global\let\svgscale\undefined%
  \makeatother%
  \begin{picture}(1,0.25518277)%
    \lineheight{1}%
    \setlength\tabcolsep{0pt}%
    \put(-0.23976261,0.35627608){\color[rgb]{0,0,0}\makebox(0,0)[lt]{\begin{minipage}{0.61054408\unitlength}\raggedright \end{minipage}}}%
    \put(0,0){\includegraphics[width=\unitlength,page=1]{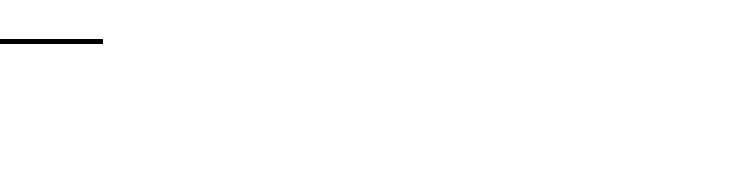}}%
    \put(0.15345259,0.21023538){\color[rgb]{0,0,0}\makebox(0,0)[lt]{\begin{minipage}{0.18561294\unitlength}\raggedright $=$\end{minipage}}}%
    \put(0,0){\includegraphics[width=\unitlength,page=2]{propagator.pdf}}%
    \put(0.35213438,0.02942054){\color[rgb]{0,0,0}\makebox(0,0)[lt]{\begin{minipage}{0.10570916\unitlength}\raggedright $+$\end{minipage}}}%
    \put(0.62554296,0.03066412){\color[rgb]{0,0,0}\makebox(0,0)[lt]{\begin{minipage}{0.10570916\unitlength}\raggedright $+$\end{minipage}}}%
    \put(0,0){\includegraphics[width=\unitlength,page=3]{propagator.pdf}}%
    \put(0.90391849,0.02962461){\color[rgb]{0,0,0}\makebox(0,0)[lt]{\begin{minipage}{0.10570916\unitlength}\raggedright $+ \ldots$\end{minipage}}}%
    \put(0.35213432,0.12056601){\color[rgb]{0,0,0}\makebox(0,0)[lt]{\begin{minipage}{0.10570916\unitlength}\raggedright $+$\end{minipage}}}%
    \put(0.62554296,0.12138921){\color[rgb]{0,0,0}\makebox(0,0)[lt]{\begin{minipage}{0.10570916\unitlength}\raggedright $+$\end{minipage}}}%
    \put(0,0){\includegraphics[width=\unitlength,page=4]{propagator.pdf}}%
    \put(0.3521343,0.21650273){\color[rgb]{0,0,0}\makebox(0,0)[lt]{\begin{minipage}{0.10570916\unitlength}\raggedright $+$\end{minipage}}}%
    \put(0.62554293,0.21650273){\color[rgb]{0,0,0}\makebox(0,0)[lt]{\begin{minipage}{0.10570916\unitlength}\raggedright $+$\end{minipage}}}%
    \put(0,0){\includegraphics[width=\unitlength,page=5]{propagator.pdf}}%
  \end{picture}%
\endgroup%

\end{equation}

This series can be organized into a Schwinger-Dyson equation,
\begin{equation}
  \label{eq:dyson_propagator}
\begingroup%
  \makeatletter%
  \providecommand\color[2][]{%
    \errmessage{(Inkscape) Color is used for the text in Inkscape, but the package 'color.sty' is not loaded}%
    \renewcommand\color[2][]{}%
  }%
  \providecommand\transparent[1]{%
    \errmessage{(Inkscape) Transparency is used (non-zero) for the text in Inkscape, but the package 'transparent.sty' is not loaded}%
    \renewcommand\transparent[1]{}%
  }%
  \providecommand\rotatebox[2]{#2}%
  \newcommand*\fsize{\dimexpr\f@size pt\relax}%
  \newcommand*\lineheight[1]{\fontsize{\fsize}{#1\fsize}\selectfont}%
  \ifx\svgwidth\undefined%
    \setlength{\unitlength}{189.08173569bp}%
    \ifx\svgscale\undefined%
      \relax%
    \else%
      \setlength{\unitlength}{\unitlength * \real{\svgscale}}%
    \fi%
  \else%
    \setlength{\unitlength}{\svgwidth}%
  \fi%
  \global\let\svgwidth\undefined%
  \global\let\svgscale\undefined%
  \makeatother%
  \begin{picture}(1,0.07679364)%
    \lineheight{1}%
    \setlength\tabcolsep{0pt}%
    \put(-0.27036742,0.19079115){\color[rgb]{0,0,0}\makebox(0,0)[lt]{\begin{minipage}{0.68847775\unitlength}\raggedright \end{minipage}}}%
    \put(0,0){\includegraphics[width=\unitlength,page=1]{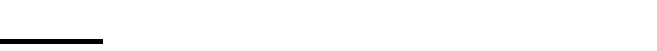}}%
    \put(0.17304024,0.0261089){\color[rgb]{0,0,0}\makebox(0,0)[lt]{\begin{minipage}{0.20930574\unitlength}\raggedright $=$\end{minipage}}}%
    \put(0,0){\includegraphics[width=\unitlength,page=2]{dyson_1PI.pdf}}%
    \put(0.39708293,0.03317626){\color[rgb]{0,0,0}\makebox(0,0)[lt]{\begin{minipage}{0.11920254\unitlength}\raggedright $+$\end{minipage}}}%
    \put(0,0){\includegraphics[width=\unitlength,page=3]{dyson_1PI.pdf}}%
    \put(0.71183162,0.03282201){\color[rgb]{0,0,0}\makebox(0,0)[lt]{\begin{minipage}{0.11920254\unitlength}\raggedright $+$\end{minipage}}}%
    \put(0,0){\includegraphics[width=\unitlength,page=4]{dyson_1PI.pdf}}%
  \end{picture}%
\endgroup%

\end{equation}
where we have introduced two kinds of one--particle irreducible diagrams (1PI) -- those with an $A$ or a $B$ as their outermost insertion. 
We remind the reader that 1PI diagrams are the amputated diagrams which cannot be disconnected by cutting a single $1/z$ line.
We solve Eq.~\eqref{eq:dyson_propagator} by introducing the `self-energy' $\Sigma_H(z)$,
\begin{align}
    \label{eq:self_energy_def} 
  f_H(z) &= \frac{1}{z - \Sigma_H(z)}
\end{align}
Comparing to Eq.~\eqref{eq:dyson_propagator}, we have
\begin{align}
\label{eq:addition_sigma} 
  \Sigma_H(z) &= \Sigma_A(z) + \Sigma_B(z)  
\end{align}
where 
\begin{align}
  \label{eq:self_energy_contributions}
\begingroup%
  \makeatletter%
  \providecommand\color[2][]{%
    \errmessage{(Inkscape) Color is used for the text in Inkscape, but the package 'color.sty' is not loaded}%
    \renewcommand\color[2][]{}%
  }%
  \providecommand\transparent[1]{%
    \errmessage{(Inkscape) Transparency is used (non-zero) for the text in Inkscape, but the package 'transparent.sty' is not loaded}%
    \renewcommand\transparent[1]{}%
  }%
  \providecommand\rotatebox[2]{#2}%
  \newcommand*\fsize{\dimexpr\f@size pt\relax}%
  \newcommand*\lineheight[1]{\fontsize{\fsize}{#1\fsize}\selectfont}%
  \ifx\svgwidth\undefined%
    \setlength{\unitlength}{186.32366991bp}%
    \ifx\svgscale\undefined%
      \relax%
    \else%
      \setlength{\unitlength}{\unitlength * \real{\svgscale}}%
    \fi%
  \else%
    \setlength{\unitlength}{\svgwidth}%
  \fi%
  \global\let\svgwidth\undefined%
  \global\let\svgscale\undefined%
  \makeatother%
  \begin{picture}(1,0.18685912)%
    \lineheight{1}%
    \setlength\tabcolsep{0pt}%
    \put(-0.55672956,0.30254414){\color[rgb]{0,0,0}\makebox(0,0)[lt]{\begin{minipage}{0.69866898\unitlength}\raggedright \end{minipage}}}%
    \put(0.64558668,0.14259614){\color[rgb]{0,0,0}\makebox(0,0)[lt]{\begin{minipage}{0.12096705\unitlength}\raggedright $+$\end{minipage}}}%
    \put(0.64558669,0.03316808){\color[rgb]{0,0,0}\makebox(0,0)[lt]{\begin{minipage}{0.12096705\unitlength}\raggedright $+$\end{minipage}}}%
    \put(0,0){\includegraphics[width=\unitlength,page=1]{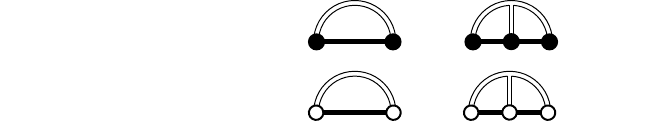}}%
    \put(0.89005016,0.1425963){\color[rgb]{0,0,0}\makebox(0,0)[lt]{\begin{minipage}{0.12096705\unitlength}\raggedright $+ \ldots$\end{minipage}}}%
    \put(0.88728888,0.03316792){\color[rgb]{0,0,0}\makebox(0,0)[lt]{\begin{minipage}{0.12096705\unitlength}\raggedright $+ \ldots$\end{minipage}}}%
    \put(0.92186971,0.12212641){\color[rgb]{0,0,0}\makebox(0,0)[lt]{\lineheight{1.25}\smash{\begin{tabular}[t]{l} \end{tabular}}}}%
    \put(-0.36347911,0.23197214){\color[rgb]{0,0,0}\makebox(0,0)[lt]{\begin{minipage}{0.62902877\unitlength}\raggedright \end{minipage}}}%
    \put(-0.00176105,0.16769966){\color[rgb]{0,0,0}\makebox(0,0)[lt]{\begin{minipage}{0.22843313\unitlength}\raggedright $\Sigma_A(z) = $\end{minipage}}}%
    \put(-0.00178201,0.05827159){\color[rgb]{0,0,0}\makebox(0,0)[lt]{\begin{minipage}{0.22843313\unitlength}\raggedright $\Sigma_B(z) = $\end{minipage}}}%
    \put(0,0){\includegraphics[width=\unitlength,page=2]{self_energies.pdf}}%
    \put(0.62067122,0.1693539){\color[rgb]{0,0,0}\makebox(0,0)[lt]{\begin{minipage}{0.10602405\unitlength}\raggedright \end{minipage}}}%
    \put(0.40407136,0.14259614){\color[rgb]{0,0,0}\makebox(0,0)[lt]{\begin{minipage}{0.12096705\unitlength}\raggedright $=$\end{minipage}}}%
    \put(0.40407137,0.03316808){\color[rgb]{0,0,0}\makebox(0,0)[lt]{\begin{minipage}{0.12096705\unitlength}\raggedright $=$\end{minipage}}}%
    \put(0,0){\includegraphics[width=\unitlength,page=3]{self_energies.pdf}}%
  \end{picture}%
\endgroup%

\end{align}

Let us pause and make several connections between planar perturbation theory and free probability theory explicit.
The self-energies in Eq.~\eqref{eq:self_energy_contributions} are given algebraically by
\begin{align}
  \label{eq:sigma_A}
  \Sigma_A(z) \equiv R_A(f(z)) \equiv \sum_{p = 0}^\infty \free{A^{n+1}} f_H(z)^n
\end{align}
and similarly for $\Sigma_B$. 
That is, the self energy $\Sigma_A$ is given by the free cumulant generating function $R_A$ evaluated at $f_H(z)$. 
Thus, additivity of the self-energy in planar perturbation theory is equivalent to the additivity of the free cumulants of freely independent random variables \cite{zeeLawAdditionRandom1996}. 

The Cartesian decomposition of $f(x)$ (eq.\ref{eq:prop_real_axis}) induces a similar structure in $\Sigma_A$ (eq.\ref{eq:sigma_A}): 
\begin{align}
  \label{eq:sigma_real_axis}
  \Sigma_A(x \pm i \delta) &= \sum_{n=1}^\infty \free{A^n} \left( \phi(x) \mp i \pi \rho(x) \right)^n \\
  &\equiv \Sigma_A^R(x) \mp i \Sigma_A^I(x)
\end{align}
where $\Sigma_A^R(x)$ and $\Sigma_A^I(x)$ are real functions.

\subsubsection{2PI Diagrams}
\label{sub:2pi_diagrams}
It is useful at this point to introduce the analog of the two--particle irreducible diagrams familiar in the context of field theory: they are (amputated) diagrams that cannot be separated into two disconnected pieces by cutting at most one $z$ and one $w$ line. Notice that all the vertices at the boundary of a 2PI diagram must be of the same type, either $A$ or $B$.

We will denote $\pi_A(z,w)$ the sum of all 2PI diagrams with the most external interaction of type $A$:
\begin{equation}
    \label{eq:2PI_diagram} 
\begingroup%
  \makeatletter%
  \providecommand\color[2][]{%
    \errmessage{(Inkscape) Color is used for the text in Inkscape, but the package 'color.sty' is not loaded}%
    \renewcommand\color[2][]{}%
  }%
  \providecommand\transparent[1]{%
    \errmessage{(Inkscape) Transparency is used (non-zero) for the text in Inkscape, but the package 'transparent.sty' is not loaded}%
    \renewcommand\transparent[1]{}%
  }%
  \providecommand\rotatebox[2]{#2}%
  \newcommand*\fsize{\dimexpr\f@size pt\relax}%
  \newcommand*\lineheight[1]{\fontsize{\fsize}{#1\fsize}\selectfont}%
  \ifx\svgwidth\undefined%
    \setlength{\unitlength}{196.228234bp}%
    \ifx\svgscale\undefined%
      \relax%
    \else%
      \setlength{\unitlength}{\unitlength * \real{\svgscale}}%
    \fi%
  \else%
    \setlength{\unitlength}{\svgwidth}%
  \fi%
  \global\let\svgwidth\undefined%
  \global\let\svgscale\undefined%
  \makeatother%
  \begin{picture}(1,0.17974543)%
    \lineheight{1}%
    \setlength\tabcolsep{0pt}%
    \put(0.41166464,0.09791974){\color[rgb]{0,0,0}\makebox(0,0)[lt]{\begin{minipage}{0.08547973\unitlength}\raggedright $=$\end{minipage}}}%
    \put(-0.00169543,0.12085339){\color[rgb]{0,0,0}\makebox(0,0)[lt]{\begin{minipage}{0.4785442\unitlength}\raggedright $\pi_A(w,z) =$\end{minipage}}}%
    \put(0,0){\includegraphics[width=\unitlength,page=1]{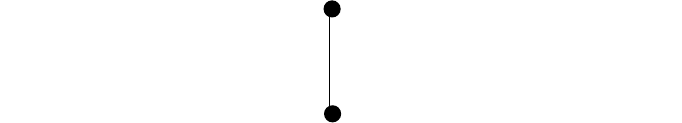}}%
    \put(0.52349118,0.10556394){\color[rgb]{0,0,0}\makebox(0,0)[lt]{\begin{minipage}{0.08547973\unitlength}\raggedright $+$\end{minipage}}}%
    \put(0.90209121,0.11320815){\color[rgb]{0,0,0}\makebox(0,0)[lt]{\begin{minipage}{0.18953057\unitlength}\raggedright $+\ldots$\end{minipage}}}%
    \put(0,0){\includegraphics[width=\unitlength,page=2]{2PI.pdf}}%
    \put(0.29778627,0.07844029){\color[rgb]{0,0,0}\makebox(0,0)[lt]{\lineheight{1.25}\smash{\begin{tabular}[t]{l}$\pi_A$\end{tabular}}}}%
    \put(0,0){\includegraphics[width=\unitlength,page=3]{2PI.pdf}}%
    \put(0.7303614,0.10556394){\color[rgb]{0,0,0}\makebox(0,0)[lt]{\begin{minipage}{0.08547973\unitlength}\raggedright $+$\end{minipage}}}%
  \end{picture}%
\endgroup%

\end{equation}
and define analogously $\pi_B(z,w)$. The series corresponding to eq.~\ref{eq:2PI_diagram} is
\begin{align}
    \pi_A(z,w) &= \sum_{n=0}^\infty \{A^{n+2}\} \sum_{t=0}^n f(z)^t f(w)^{n-t} \\
    &= \frac{ \Sigma_A(z) - \Sigma_A(w)}{f(z) - f(w)} \label{eq:2PI_sigma} 
\end{align}

The linearity of the self--energy contributions guarantees the linearity of this quantity as well: 
\begin{equation}
    \pi_H(z,w) = \pi_A(z,w) + \pi_B(z,w)
\end{equation}

\subsubsection{Box Resummation}
\label{sub:box_diagram}
In the diagrammatic expansion of $G(z,w)$ we need a set of disconnected diagrams 
that is closely related to the 2PI we just introduced: they are
composed of a $1/z$ line, a $1/w$ line and any number of
vertices possibly connecting those two lines. We call the sum of this class of
diagrams the \emph{box}, and represent it with a gray shaded area.
\footnote{The actual shape does not matter, as long as it has four sides: one
side must contain only $1/z$ lines and the opposite one only contains $1/w$
lines} The resummation of these diagrams proceeds in a way analogous to what we did for the propagator.

Before writing the self--consistent equation we show a few of
the diagrams that contribute to the box:
\begin{equation}
\begingroup%
  \makeatletter%
  \providecommand\color[2][]{%
    \errmessage{(Inkscape) Color is used for the text in Inkscape, but the package 'color.sty' is not loaded}%
    \renewcommand\color[2][]{}%
  }%
  \providecommand\transparent[1]{%
    \errmessage{(Inkscape) Transparency is used (non-zero) for the text in Inkscape, but the package 'transparent.sty' is not loaded}%
    \renewcommand\transparent[1]{}%
  }%
  \providecommand\rotatebox[2]{#2}%
  \newcommand*\fsize{\dimexpr\f@size pt\relax}%
  \newcommand*\lineheight[1]{\fontsize{\fsize}{#1\fsize}\selectfont}%
  \ifx\svgwidth\undefined%
    \setlength{\unitlength}{224.62732991bp}%
    \ifx\svgscale\undefined%
      \relax%
    \else%
      \setlength{\unitlength}{\unitlength * \real{\svgscale}}%
    \fi%
  \else%
    \setlength{\unitlength}{\svgwidth}%
  \fi%
  \global\let\svgwidth\undefined%
  \global\let\svgscale\undefined%
  \makeatother%
  \begin{picture}(1,0.31769847)%
    \lineheight{1}%
    \setlength\tabcolsep{0pt}%
    \put(0.31939315,0.25610858){\color[rgb]{0,0,0}\makebox(0,0)[lt]{\begin{minipage}{0.11516221\unitlength}\raggedright $=$\end{minipage}}}%
    \put(-0.00295922,0.27237819){\color[rgb]{0,0,0}\makebox(0,0)[lt]{\begin{minipage}{0.418043\unitlength}\raggedright $h(w,z) =$\end{minipage}}}%
    \put(0,0){\includegraphics[width=\unitlength,page=1]{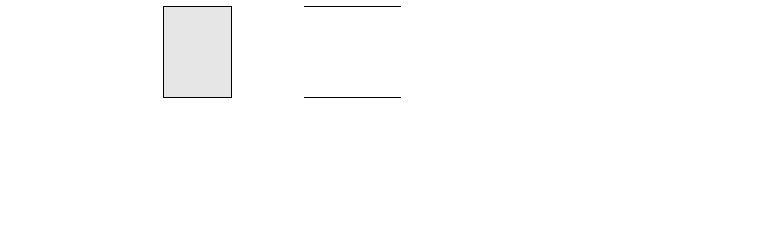}}%
    \put(0.53120951,0.25610871){\color[rgb]{0,0,0}\makebox(0,0)[lt]{\begin{minipage}{0.11587149\unitlength}\raggedright $+$\end{minipage}}}%
    \put(0,0){\includegraphics[width=\unitlength,page=2]{boxdiagram.pdf}}%
    \put(0.73632565,0.25610871){\color[rgb]{0,0,0}\makebox(0,0)[lt]{\begin{minipage}{0.12914069\unitlength}\raggedright $+$\end{minipage}}}%
    \put(0,0){\includegraphics[width=\unitlength,page=3]{boxdiagram.pdf}}%
    \put(0.73593603,0.08782898){\color[rgb]{0,0,0}\makebox(0,0)[lt]{\begin{minipage}{0.33853766\unitlength}\raggedright $+\quad \ldots$\end{minipage}}}%
    \put(0,0){\includegraphics[width=\unitlength,page=4]{boxdiagram.pdf}}%
    \put(0.31939303,0.08850398){\color[rgb]{0,0,0}\makebox(0,0)[lt]{\begin{minipage}{0.1148415\unitlength}\raggedright $+$\end{minipage}}}%
    \put(0.53120941,0.08850398){\color[rgb]{0,0,0}\makebox(0,0)[lt]{\begin{minipage}{0.1431716\unitlength}\raggedright $+$\end{minipage}}}%
    \put(0,0){\includegraphics[width=\unitlength,page=5]{boxdiagram.pdf}}%
  \end{picture}%
\endgroup%

\end{equation}
where we can imagine that the first line contains all the disconnected diagrams and the second line all the connected ones.

The disconnected diagrams can be summed using the results of
sec.~\ref{sub:propagator_renormalization},  while the connected diagrams can be written in terms of $\pi_H$ and the box itself: 
\begin{equation}
\begingroup%
  \makeatletter%
  \providecommand\color[2][]{%
    \errmessage{(Inkscape) Color is used for the text in Inkscape, but the package 'color.sty' is not loaded}%
    \renewcommand\color[2][]{}%
  }%
  \providecommand\transparent[1]{%
    \errmessage{(Inkscape) Transparency is used (non-zero) for the text in Inkscape, but the package 'transparent.sty' is not loaded}%
    \renewcommand\transparent[1]{}%
  }%
  \providecommand\rotatebox[2]{#2}%
  \newcommand*\fsize{\dimexpr\f@size pt\relax}%
  \newcommand*\lineheight[1]{\fontsize{\fsize}{#1\fsize}\selectfont}%
  \ifx\svgwidth\undefined%
    \setlength{\unitlength}{161.47613165bp}%
    \ifx\svgscale\undefined%
      \relax%
    \else%
      \setlength{\unitlength}{\unitlength * \real{\svgscale}}%
    \fi%
  \else%
    \setlength{\unitlength}{\svgwidth}%
  \fi%
  \global\let\svgwidth\undefined%
  \global\let\svgscale\undefined%
  \makeatother%
  \begin{picture}(1,0.18975991)%
    \lineheight{1}%
    \setlength\tabcolsep{0pt}%
    \put(0.41665687,0.10491731){\color[rgb]{0,0,0}\makebox(0,0)[lt]{\begin{minipage}{0.10387626\unitlength}\raggedright $=$\end{minipage}}}%
    \put(-0.00206031,0.13278663){\color[rgb]{0,0,0}\makebox(0,0)[lt]{\begin{minipage}{0.58153414\unitlength}\raggedright $h(w,z) =$\end{minipage}}}%
    \put(0.60623431,0.11420649){\color[rgb]{0,0,0}\makebox(0,0)[lt]{\begin{minipage}{0.10387626\unitlength}\raggedright $+$\end{minipage}}}%
    \put(0,0){\includegraphics[width=\unitlength,page=1]{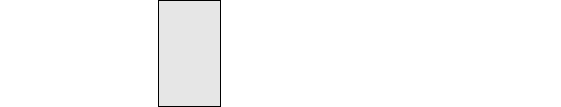}}%
    \put(0.79493839,0.08182605){\color[rgb]{0,0,0}\makebox(0,0)[lt]{\lineheight{1.25}\smash{\begin{tabular}[t]{l}$\pi_H$\end{tabular}}}}%
    \put(0.22294317,0.81169865){\color[rgb]{0,0,0}\makebox(0,0)[lt]{\begin{minipage}{0.29609639\unitlength}\raggedright \end{minipage}}}%
    \put(0,0){\includegraphics[width=\unitlength,page=2]{box_2PI.pdf}}%
    \put(0.57638814,0.29640466){\color[rgb]{0,0,0}\makebox(0,0)[lt]{\begin{minipage}{0.50167112\unitlength}\raggedright \end{minipage}}}%
    \put(0.95325735,0.28901506){\color[rgb]{0,0,0}\makebox(0,0)[lt]{\begin{minipage}{0.6223679\unitlength}\raggedright \end{minipage}}}%
  \end{picture}%
\endgroup%

\end{equation}
which can be solved for $h(z,w)$ to give
\begin{equation}
h(z,w) = \frac{f(z)f(w)}{1 - \pi_H(z,w)f(z)f(w)}
\end{equation}

We can further simplify this expression using eq.~\ref{eq:2PI_sigma} and eq.~\ref{eq:self_energy_def}, so that after some algebra we obtain
\begin{equation}
  \label{eq:nongaussian_box}
  h(z,w) = - \frac{f(z) - f(w)}{z - w}
\end{equation}

\subsubsection{Triangle Resummation}
\label{sub:triangle_resummation}
There is another pattern that is important for the computation of the two--point function: (amputated) diagrams in which one of the two special insertions of $A$ participates in a vertex with at least three legs, and it connects to both a $z$ and a $w$ line. The sum of all diagrams of this kind is
\begin{equation}
\begingroup%
  \makeatletter%
  \providecommand\color[2][]{%
    \errmessage{(Inkscape) Color is used for the text in Inkscape, but the package 'color.sty' is not loaded}%
    \renewcommand\color[2][]{}%
  }%
  \providecommand\transparent[1]{%
    \errmessage{(Inkscape) Transparency is used (non-zero) for the text in Inkscape, but the package 'transparent.sty' is not loaded}%
    \renewcommand\transparent[1]{}%
  }%
  \providecommand\rotatebox[2]{#2}%
  \newcommand*\fsize{\dimexpr\f@size pt\relax}%
  \newcommand*\lineheight[1]{\fontsize{\fsize}{#1\fsize}\selectfont}%
  \ifx\svgwidth\undefined%
    \setlength{\unitlength}{177.37669373bp}%
    \ifx\svgscale\undefined%
      \relax%
    \else%
      \setlength{\unitlength}{\unitlength * \real{\svgscale}}%
    \fi%
  \else%
    \setlength{\unitlength}{\svgwidth}%
  \fi%
  \global\let\svgwidth\undefined%
  \global\let\svgscale\undefined%
  \makeatother%
  \begin{picture}(1,0.18915358)%
    \lineheight{1}%
    \setlength\tabcolsep{0pt}%
    \put(0,0){\includegraphics[width=\unitlength,page=1]{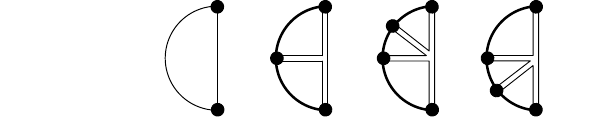}}%
    \put(-0.01050861,0.13066676){\color[rgb]{0,0,0}\makebox(0,0)[lt]{\begin{minipage}{0.17522518\unitlength}\raggedright \end{minipage}}}%
    \put(-0.0037438,0.0799888){\color[rgb]{0,0,0}\makebox(0,0)[lt]{\lineheight{1.25}\smash{\begin{tabular}[t]{l}$T(z,w) =$\end{tabular}}}}%
    \put(-0.00231963,0.14543401){\color[rgb]{0,0,0}\makebox(0,0)[lt]{\begin{minipage}{0.18264998\unitlength}\raggedright \end{minipage}}}%
    \put(0.14805508,0.21980217){\color[rgb]{0,0,0}\makebox(0,0)[lt]{\begin{minipage}{0.70090074\unitlength}\raggedright \end{minipage}}}%
    \put(0.37103697,0.1029827){\color[rgb]{0,0,0}\makebox(0,0)[lt]{\begin{minipage}{0.26988706\unitlength}\raggedright $=$\end{minipage}}}%
    \put(0.54862512,0.11143928){\color[rgb]{0,0,0}\makebox(0,0)[lt]{\begin{minipage}{0.10468318\unitlength}\raggedright $+$\end{minipage}}}%
    \put(0.71493403,0.23638467){\color[rgb]{0,0,0}\makebox(0,0)[lt]{\begin{minipage}{0.15737094\unitlength}\raggedright \end{minipage}}}%
    \put(0.71775669,0.11143928){\color[rgb]{0,0,0}\makebox(0,0)[lt]{\begin{minipage}{0.10468318\unitlength}\raggedright $+$\end{minipage}}}%
    \put(0.29533868,0.07998881){\color[rgb]{0,0,0}\makebox(0,0)[lt]{\lineheight{1.25}\smash{\begin{tabular}[t]{l}$T$\end{tabular}}}}%
    \put(0.88298634,0.07153223){\color[rgb]{0,0,0}\makebox(0,0)[lt]{\lineheight{1.25}\smash{\begin{tabular}[t]{l}$+ \ldots$\end{tabular}}}}%
  \end{picture}%
\endgroup%

\end{equation}
which can be written in a simple form using eq.~\ref{eq:sigma_A}:
\begin{align}
    T(z,w) &= \sum_{ n = 0}^\infty \{ A^{n+3} \} f(z) f(w) \sum_{ t = 0}^n f(z)^t f(w)^{n-t} \\
    &= \frac{f(w) \Sigma_A(z) - f(z) \Sigma_A(w)}{f(z) - f(w)}
\end{align}

As a simple consistency check, notice that if $A$ is from one of the Gaussian ensembles we have $\Sigma_A(z) = \{ A^2 \} f(z)$ and thus $T(z,w)$ vanishes identically. This is consistent with the diagrammatic statement that Gaussian matrices do not have vertices with more than two legs.

\subsection{Two--point correlation function}
We are now ready to compute the two--points function $G(z,w)$: as
stated in sec.~\ref{sub:diagrammatic_rules}, we must sum over all
the circular planar diagrams with at least two $A$ insertions.

Using the objects $\Sigma_A$, $h$, and $T$ we can sum all diagrams in which the two special $A$ insertions do not connect to each other. 

We organize the remaining diagrams by the kind of interaction the two $A$ participate in:
\begin{equation}
\begingroup%
  \makeatletter%
  \providecommand\color[2][]{%
    \errmessage{(Inkscape) Color is used for the text in Inkscape, but the package 'color.sty' is not loaded}%
    \renewcommand\color[2][]{}%
  }%
  \providecommand\transparent[1]{%
    \errmessage{(Inkscape) Transparency is used (non-zero) for the text in Inkscape, but the package 'transparent.sty' is not loaded}%
    \renewcommand\transparent[1]{}%
  }%
  \providecommand\rotatebox[2]{#2}%
  \newcommand*\fsize{\dimexpr\f@size pt\relax}%
  \newcommand*\lineheight[1]{\fontsize{\fsize}{#1\fsize}\selectfont}%
  \ifx\svgwidth\undefined%
    \setlength{\unitlength}{220.11610136bp}%
    \ifx\svgscale\undefined%
      \relax%
    \else%
      \setlength{\unitlength}{\unitlength * \real{\svgscale}}%
    \fi%
  \else%
    \setlength{\unitlength}{\svgwidth}%
  \fi%
  \global\let\svgwidth\undefined%
  \global\let\svgscale\undefined%
  \makeatother%
  \begin{picture}(1,0.48856034)%
    \lineheight{1}%
    \setlength\tabcolsep{0pt}%
    \put(-0.00301459,0.44134281){\color[rgb]{0,0,0}\makebox(0,0)[lt]{\begin{minipage}{0.32043019\unitlength}\raggedright $G(z,w)=$\end{minipage}}}%
    \put(0.76722507,0.43453216){\color[rgb]{0,0,0}\makebox(0,0)[lt]{\begin{minipage}{0.1246181\unitlength}\raggedright $+$\end{minipage}}}%
    \put(0.61072758,0.08313488){\color[rgb]{0,0,0}\makebox(0,0)[lt]{\begin{minipage}{0.33342178\unitlength}\raggedright $+ \ldots$\end{minipage}}}%
    \put(0,0){\includegraphics[width=\unitlength,page=1]{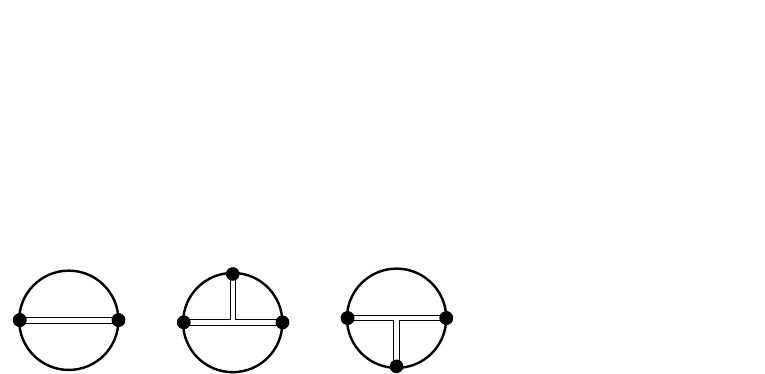}}%
    \put(0.36419409,0.43453217){\color[rgb]{0,0,0}\makebox(0,0)[lt]{\begin{minipage}{0.1246181\unitlength}\raggedright $+$\end{minipage}}}%
    \put(0.56339021,0.43472294){\color[rgb]{0,0,0}\makebox(0,0)[lt]{\begin{minipage}{0.1246181\unitlength}\raggedright $+$\end{minipage}}}%
    \put(0.18020031,0.08313488){\color[rgb]{0,0,0}\makebox(0,0)[lt]{\begin{minipage}{0.1246181\unitlength}\raggedright $+$\end{minipage}}}%
    \put(0.39302561,0.08313488){\color[rgb]{0,0,0}\makebox(0,0)[lt]{\begin{minipage}{0.1246181\unitlength}\raggedright $+$\end{minipage}}}%
    \put(0,0){\includegraphics[width=\unitlength,page=2]{non_gaussian_corr.pdf}}%
    \put(-0.1407109,0.78091137){\color[rgb]{0,0,0}\makebox(0,0)[lt]{\begin{minipage}{2.05130304\unitlength}\raggedright \end{minipage}}}%
    \put(-0.0364035,0.08313488){\color[rgb]{0,0,0}\makebox(0,0)[lt]{\begin{minipage}{0.1246181\unitlength}\raggedright $+$\end{minipage}}}%
    \put(0.76722507,0.25735294){\color[rgb]{0,0,0}\makebox(0,0)[lt]{\begin{minipage}{0.1246181\unitlength}\raggedright $+$\end{minipage}}}%
    \put(0.36419409,0.25735295){\color[rgb]{0,0,0}\makebox(0,0)[lt]{\begin{minipage}{0.1246181\unitlength}\raggedright $+$\end{minipage}}}%
    \put(0.56339021,0.25754372){\color[rgb]{0,0,0}\makebox(0,0)[lt]{\begin{minipage}{0.1246181\unitlength}\raggedright $+$\end{minipage}}}%
    \put(0.16231199,0.25777886){\color[rgb]{0,0,0}\makebox(0,0)[lt]{\begin{minipage}{0.1246181\unitlength}\raggedright $+$\end{minipage}}}%
    \put(-0.03448689,0.25797158){\color[rgb]{0,0,0}\makebox(0,0)[lt]{\begin{minipage}{0.1246181\unitlength}\raggedright $+$\end{minipage}}}%
    \put(0,0){\includegraphics[width=\unitlength,page=3]{non_gaussian_corr.pdf}}%
    \put(0.23418077,0.23558417){\color[rgb]{0,0,0}\makebox(0,0)[lt]{\lineheight{1.25}\smash{\begin{tabular}[t]{l}$T$\end{tabular}}}}%
    \put(0,0){\includegraphics[width=\unitlength,page=4]{non_gaussian_corr.pdf}}%
    \put(0.43506289,0.23558417){\color[rgb]{0,0,0}\makebox(0,0)[lt]{\lineheight{1.25}\smash{\begin{tabular}[t]{l}$T$\end{tabular}}}}%
    \put(0,0){\includegraphics[width=\unitlength,page=5]{non_gaussian_corr.pdf}}%
    \put(0.70234376,0.23558416){\color[rgb]{0,0,0}\makebox(0,0)[lt]{\lineheight{1.25}\smash{\begin{tabular}[t]{l}$T$\end{tabular}}}}%
    \put(0,0){\includegraphics[width=\unitlength,page=6]{non_gaussian_corr.pdf}}%
    \put(0.90457093,0.23558428){\color[rgb]{0,0,0}\makebox(0,0)[lt]{\lineheight{1.25}\smash{\begin{tabular}[t]{l}$T$\end{tabular}}}}%
    \put(0,0){\includegraphics[width=\unitlength,page=7]{non_gaussian_corr.pdf}}%
    \put(0.02482083,0.23558301){\color[rgb]{0,0,0}\makebox(0,0)[lt]{\lineheight{1.25}\smash{\begin{tabular}[t]{l}$T$\end{tabular}}}}%
    \put(0.11341044,0.23558301){\color[rgb]{0,0,0}\makebox(0,0)[lt]{\lineheight{1.25}\smash{\begin{tabular}[t]{l}$T$\end{tabular}}}}%
  \end{picture}%
\endgroup%

\end{equation}
which is the diagrammatic representation of
\begin{equation}
  \begin{split}
  G(z,w) &= h(z,w) \left( T(z,w) + \Sigma_A(z) + \Sigma_A(w) \right)^2 \\
  &+ \sum_{n=0}^\infty \free{A^{n+2}} f(z) f(w) \sum_{t=0}^{n} f(z)^{t} f(w)^{n-t}
  \end{split}
\end{equation}
and the series appearing in the last term can be summed as usual using eq.~\ref{eq:sigma_A}:
\begin{equation}
  \label{eq:ng_freq_corr}
\boxed{
  \begin{split}
  G(w,z) &= f(z)f(w) \frac{ \Sigma_A(w) - \Sigma_A(z)}{f(w) - f(z)} \\
  & - \frac{f(z) - f(w)}{z - w}  \left( \frac{f(z) \Sigma_A(z) - f(w) \Sigma_A(w)}{f(z) - f(w)} \right)^2
  \end{split}}
\end{equation}

This is our main result about the two--point function:
once we fix the probability distributions
of $A$ and $B$, we can compute
\cite{novakThreeLecturesFree2012,zeeLawAdditionRandom1996} $f(z)$ and
$\Sigma_A(z)$, so they can be considered inputs to the problem.

\subsection{Example: $A,B$ Gaussian}
\label{sub:gaussian_correlator}
If $A$ and $B$ are sampled from the GUE ensemble, all their free cumulants vanish after the second one. Without loss of generality we can set 
\begin{equation}
    \{ A \} = \{ B \} = 0,
\end{equation}
while a convenient choice for the second free cumulant is 
\begin{equation}
  \label{eq:free_cumul_gauss}
  \{ A^2 \} = \lambda, \quad \{ B^2 \} = 1 - \lambda
\end{equation}
with $0 \leq \lambda \leq 1$.
For these random variables the self--energy series terminates after just one term:
\begin{equation}
  \Sigma_A(z) = \lambda f(z) \qquad 
  \Sigma_B(z) = (1-\lambda) f(z).
\end{equation}

In the infinite $N$ limit, independent GUE variables are also freely independent \cite{nicaLecturesCombinatoricsFree2006}, so we can use eq.~\ref{eq:addition_sigma} and eq.~\ref{eq:self_energy_def} to compute the propagator for $H$:
\begin{equation}
    \label{eq:gue_propagator} 
  f(z) = \frac{1}{2} \left(z - \sqrt{z^2 - 4}\right)
\end{equation}
and plugging $f$ and $\Sigma_A$ in eq.~\ref{eq:ng_freq_corr} we have
\begin{equation}
\begin{split}
  G(w,z) = &\lambda f(z)f(w) \\
  &-\lambda^2 \frac{f(z) - f(w)}{z - w}  \left(f(w) + f(z)\right)^2.
\end{split}
\end{equation}

The real time correlator is (eq.~\ref{eq:inverse_cauchy}):
\begin{equation}
  \avg{A(t)A(0)} = \oint \frac{dz}{2 \pi i} \frac{dw}{2 \pi i} e^{i(z-w)t} G(z,w)
\end{equation}
which has a closed form 
\footnote{The factor $1 / \lambda$ on the left--hand side is included for convenience: with the choices of this section
  $\avg{A(0)A(0)}/ \lambda =  1$.}
in terms of the Bessel function~$J_1$:
\begin{equation}
  \label{eq:gaussian_time_corr}
  \frac{1}{\lambda} \avg{A(t)A(0)} = \lambda + (1- \lambda) \left(\frac{J_1(2t)}{t}\right)^2
\end{equation}

\section{Late Time Asymptotics}
\label{sec:asymptotics}

\subsection{General Case} 
\label{sub:general_case}

The integral in eq.~\ref{eq:inverse_cauchy} can not be expressed in terms of elementary functions except in a few special cases. In this section we prove that for a large class of Haar--invariant ensembles we have the asymptotic result
\begin{equation}
    \label{eq:long_time_form_gauss}
  \avg{A(t) A(0)} \sim  A_\infty^2  + C_{A,H} \abs{\avg{e^{i H t}}}^2 \quad t \to \infty
\end{equation}
where $C_{A,H}$ is a real constant and
\begin{equation}
     A_\infty^2 \equiv \int dx \ \rho_H(x) \bigg(\text{Re} \Sigma_A(x) + \text{Re}f_H(x) \frac{\text{Im}\Sigma_A(x)}{\pi \rho_H(x)}  \bigg)^2
\end{equation}
This expression lets us compute the long--time value of the correlator, and contains the non--trivial result that the decay is controlled by the Hamiltonian only. 

To derive eq.~\ref{eq:long_time_form_gauss}, we deform the integration contours in eq.~\ref{eq:inverse_cauchy} to run infinitesimally close to the real axis:
\begin{equation}
  z \eqqcolon x \pm i \delta \qquad w \eqqcolon y \pm i \delta \qquad x,y,\delta \in \mathbb{R}
\end{equation}
and for convenience we define the symbol
\begin{equation}
  \Delta_x^\delta f(x) \coloneqq \lim_{\delta \to 0^+} (f(x+i\delta) - f(x-i\delta))
\end{equation}
so that after parametrization the integral representation of the two--point correlator becomes
\begin{equation}
    \label{eq:int_repr_real}
     \avg{A(t)A(0)} = \int_{-\infty}^\infty \frac{dx dy}{(2 \pi i)^2} \ e^{i(x-y)t} 
\Delta_x^\delta \Delta_y^\delta G(x,y).
\end{equation}

Comparing with Eq.~\eqref{eq:ng_freq_corr}, it is clear that contributions to the long--time value of this integral only come from Dirac $\delta(x)$ terms in $f(z)$, and the divergence in $h(x + i \delta, y - i \delta)$ as $x$ approaches $y$ (see Eq.\eqref{eq:nongaussian_box}) .

The first kind of constant is trivial, so we will assume that
\begin{equation}
  \lim_{t \to \infty} \avg{e^{i H t}} = 0
\end{equation}
and focus on the second kind.

The asymptotic behavior of the inverse Fourier transform $\mathcal{F}^{-1}\{\circ\}(t)$ is determined by singularities in frequency space\cite{lighthillIntroductionFourierAnalysis1958}, with smaller positive powers $ \alpha_i > 0$ corresponding to slower real time decay:
\begin{equation}
\mathcal{F}^{-1} \bigg\{ \sum_i \abs{x - x_{i}}^{\alpha_i} \bigg\} \sim O \left( \frac{1}{\abs{t}^{\text{min}_i \alpha_i + 1}} \right) \quad t \to \infty.
\end{equation}

For many common choices of $A$ and $B$, including the Gaussian, the Orthogonal Polynomial\cite{brezinPlanarDiagrams1993}, and the Wishart ensemble, all the singularities in $ \rho_H(x)$ are of the form 
\begin{equation}
\label{eq:dos_singularity} 
\abs{x - x_i}^{\alpha_i} \theta(x-x_i) \qquad \alpha_i > 0,
\end{equation}
so we obtain the asymptotics of eq.~\ref{eq:int_repr_real} expanding the integrand in powers of $\rho$ and integrating term by term. The expansion coefficients depend on $x$ and $y$, so they could in principle modify the singular behavior, but in hindsight we realize that this not the case.

Keeping terms up to $O( \rho(x) \rho(y))$ gives
\begin{widetext}
\begin{equation}
\begin{split}
    \label{eq:Gxy_approx}
   &\Delta_x^\delta \Delta_y^\delta G(x,y) \sim (2 \pi i)^2  \bigg( R_A(\phi(x)) + \phi(x) \frac{d R_A(\phi(x))} {d \phi(x)}\bigg)^2 \delta(x-y) \rho(x) \\
   + &(2 \pi i)^2  \frac{d^2}{d \phi(x) d \phi(y)} \Bigg( 
   \phi(x) \phi(y) \frac{R_A(\phi(x)) - R_A(\phi(y))}{\phi(x) - \phi(y)} 
   - \frac{\phi(x) - \phi(y)}{x - y} \bigg(\frac{\phi(x)R_A(\phi(x)) - \phi(y)R_A(\phi(y))}{\phi(x) - \phi(y)}\bigg)^2 \Bigg) \rho(x) \rho(y).
\end{split}
\end{equation}
\end{widetext}
The first term is time--independent after integration, and gives an approximate value of $\avg{A(\infty)A(0)}$, but we can do better: using eq.~\ref{eq:sp} \emph{before} expanding in powers of $ \rho$ and collecting all terms proportional to $ \delta(x-y)$, we obtain the exact expression
\begin{equation}
\label{eq:constant} 
 \avg{A(\infty)A(0)} = \int dx \rho_H(x) \bigg(\text{Re} \Sigma_A(x) + \phi_H(x) \frac{\text{Im}\Sigma_A(x)}{\pi \rho_H(x)}  \bigg)^2  
\end{equation}
where we have reintroduced the subscript in $\rho_H(x)$ to make clear that it is the spectral density of $H$, not~$A$.

\subsection{Example: Asymptotics for a $\sqrt{x}$ Edge}
\label{sub:asymptotic_analysic_for_gue}
The square--root singularity at the edges of the GUE spectral density is very common, so we believe it is useful to analyze in detail this case. Using the same conventions as in section \ref{sub:gaussian_correlator}, eq.~\ref{eq:inverse_cauchy} reads
\begin{equation}
  \label{eq:integral_for_asymptotics}
  \begin{split}
  \frac{\avg{A(t)A(0)}}{\lambda}  = \oint & \frac{dz}{2 \pi i} \frac{dw}{2 \pi i} e^{i(z-w)t} \bigg(f(z)f(w) \\
  &-\lambda \frac{f(z) - f(w)}{z - w}  \left(f(w) + f(z)\right)^2\bigg)
  \end{split}
\end{equation}

\begin{figure}
\centering
\begingroup%
  \makeatletter%
  \providecommand\color[2][]{%
    \errmessage{(Inkscape) Color is used for the text in Inkscape, but the package 'color.sty' is not loaded}%
    \renewcommand\color[2][]{}%
  }%
  \providecommand\transparent[1]{%
    \errmessage{(Inkscape) Transparency is used (non-zero) for the text in Inkscape, but the package 'transparent.sty' is not loaded}%
    \renewcommand\transparent[1]{}%
  }%
  \providecommand\rotatebox[2]{#2}%
  \newcommand*\fsize{\dimexpr\f@size pt\relax}%
  \newcommand*\lineheight[1]{\fontsize{\fsize}{#1\fsize}\selectfont}%
  \ifx\svgwidth\undefined%
    \setlength{\unitlength}{211.317242bp}%
    \ifx\svgscale\undefined%
      \relax%
    \else%
      \setlength{\unitlength}{\unitlength * \real{\svgscale}}%
    \fi%
  \else%
    \setlength{\unitlength}{\svgwidth}%
  \fi%
  \global\let\svgwidth\undefined%
  \global\let\svgscale\undefined%
  \makeatother%
  \begin{picture}(1,0.31774068)%
    \lineheight{1}%
    \setlength\tabcolsep{0pt}%
    \put(0.37289462,0.1218043){\color[rgb]{0,0,0}\makebox(0,0)[lt]{\begin{minipage}{0.36569035\unitlength}\raggedright Re$(z)$\end{minipage}}}%
    \put(0.13308477,0.32333774){\color[rgb]{0,0,0}\makebox(0,0)[lt]{\begin{minipage}{0.36569035\unitlength}\raggedright Im$(z)$\end{minipage}}}%
    \put(0.93712087,0.11929049){\color[rgb]{0,0,0}\makebox(0,0)[lt]{\begin{minipage}{0.36569035\unitlength}\raggedright Re$(z)$\end{minipage}}}%
    \put(0.67193247,0.32365364){\color[rgb]{0,0,0}\makebox(0,0)[lt]{\begin{minipage}{0.36569035\unitlength}\raggedright Im$(z)$\end{minipage}}}%
    \put(0,0){\includegraphics[width=\unitlength,page=1]{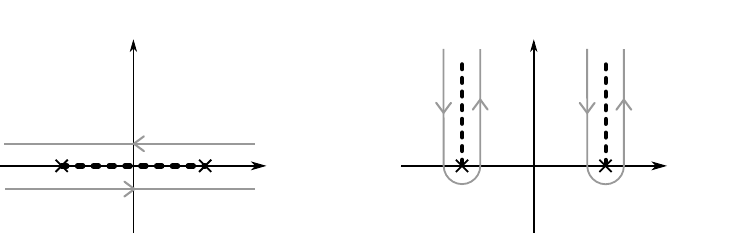}}%
    \put(0.61293402,0.03693422){\color[rgb]{0,0,0}\makebox(0,0)[lt]{\lineheight{1.25}\smash{\begin{tabular}[t]{l}$\gamma_L$\end{tabular}}}}%
    \put(0.8118189,0.03693422){\color[rgb]{0,0,0}\makebox(0,0)[lt]{\lineheight{1.25}\smash{\begin{tabular}[t]{l}$\gamma_R$\end{tabular}}}}%
  \end{picture}%
\endgroup%

\caption{\label{fig:fourier_contour} Branch cut (dashed line) and integration contour for eq.~\ref{eq:integral_for_asymptotics}. The same choices are made in the $w$ plane.}
\end{figure}
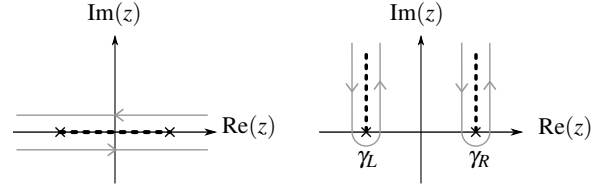

The integrand has four branching points in $z = \pm 2$ and $w = \pm 2$ (see eq.~\ref{eq:gue_propagator}), so we cut the complex plane and deform the integration contour as presented in fig.~\ref{fig:fourier_contour}.

The first term in eq.~\ref{eq:integral_for_asymptotics} is factorized and can be estimated  using standard
\footnote{We remind the reader that  $\sqrt{-1} = \exp{(-i \pi/2)}$ with our choice of branch cut.}
methods:
\begin{align}
\int_{\gamma_L} \frac{dz}{2 \pi i} \ e^{i z t} f(z)
&\sim  - \frac{e^{-i \pi/4}}{2 \sqrt{\pi}} \frac{e^{-2it}}{t^{3/2}}  &t \to \infty \\
  \int_{\gamma_R} \frac{dz}{2 \pi i} \ e^{i z t} f(z) & \sim - \frac{e^{i \pi/4}}{2 \sqrt{\pi}} \frac{e^{2it}}{t^{3/2}} &t \to \infty
\end{align}
which gives
\begin{equation}
  \label{eq:gauss_prop_asym}
  \oint  dz \ e^{i z t} f(z) \sim - \frac{1}{\sqrt{\pi}} \frac{1}{t^{3/2}} \cos \left(2t + \frac{\pi}{4} \right) \qquad t \to \infty
\end{equation}

Using eq.~\ref{eq:gue_propagator}, we can rewrite the remaining terms in eq.~\ref{eq:integral_for_asymptotics} as
\begin{equation}
  \begin{split}
  \frac{f(z) - f(w)}{z - w} & \left(f(w) + f(z)\right)^2 = \\
  & -1 + f(z)f(w) - \frac{z^2 f(z) - w^2 f(w)}{z -w}
  \end{split}
\end{equation}
During the calculation we can drop all analytic terms from the sum, as they integrate to zero on a closed contour.
The factorized term gives the same result as eq.~\ref{eq:gauss_prop_asym}, and one can prove that
\begin{equation}
  \oint \frac{dz}{2 \pi i} \frac{dw}{2 \pi i} \ e^{i(z-w)t} \ \frac{z^2 f(z) - w^2 f(w)}{w -z} = 1
\end{equation}
Even without going into the details of integration, we
can recognize that it must be constant, by taking the time derivative under the
integral sign:
\begin{align}
  \frac{d}{dt} &\oint \frac{dz}{2 \pi i} \frac{dw}{2 \pi i} e^{i(z-w)t} \frac{z^2 f(z) - w^2 f(w)}{w -z} \\
  = &\oint \frac{dz}{2 \pi i} \frac{dw}{2 \pi i} e^{i(z-w)t} i (z^2 f(z) - w^2 f(w)) = 0
\end{align}
In conclusion, we have proved that
\begin{equation}
  \frac{\avg{A(t)A(0)}}{\lambda} \sim \lambda + \frac{1 - \lambda}{\pi t^3} \cos^2\left(2 t + \frac{\pi}{4}\right) \quad t \to \infty
\end{equation}
which is consistent with the exact result eq.\ref{eq:gaussian_time_corr}, since 
\begin{equation}
  \frac{J_1(2t)}{t} \sim - \frac{1}{ \sqrt{ \pi}} \frac{1}{t^{3/2}} \cos\left(2 t + \frac{\pi}{4}\right)\qquad t \to \infty
\end{equation}

\section{Finite temperature modifications}
\label{sec:finite_temperature}

In order to shift from infinite to finite temperature $\beta^{-1}$, observe that the exact analysis of $G(w,z)$ is unmodified, so the results we have derived about $G(t)$ immediately transfer \emph{mutatis mutandis}: it is sufficient to replace $t \to t+i\beta$ in appropriate locations, which  induces some thermal reweighting of the integrals. Here, we point out the key modifications that need to be made in  Secs.~\ref{sec:exact_analysis} and \ref{sec:asymptotics} for finite $\beta$.

At finite temperature, the dynamical autocorrelator is
\begin{align}
    \langle A(t) A(0) \rangle_\beta = \mathbb{E} \left[ \frac{\Tr{e^{-\beta H} A(t) A(0)}}{\Tr{e^{-\beta H}}} \right]
\end{align}
At large $N$, the concentration of measure allows one to split the disorder average between numerator and denominator. Defining the partition function
\begin{equation}
    Z \equiv \frac{1}{N} \mathbb{E} \left[ \Tr{e^{-\beta H}} \right]
\end{equation}
the correlator becomes
\begin{align}
    \langle A(t) A(0) \rangle_\beta = \frac{1}{Z} \langle e^{(it - \beta)H} A e^{-i H t} A \rangle
\end{align}
where the average on the right is the usual disordered averaged trace from Eq.~\eqref{eq:free_prob_functional}.

Retracing the steps of Sec.\ref{sec:exact_analysis}, we find that all the information about temperature disappears from $G(z,w)$, and is only contained in the integration measure:
\begin{equation}
  \label{eq:finite_temperature_cauchy_transform}
   \langle A(t) A(0) \rangle_\beta = \oint \frac{dw}{2 \pi i} \frac{dz}{2 \pi i} \frac{e^{(it-\beta)z}}{Z} e^{-iwt} G(w,z)
\end{equation}
The exponential factor $e^{-\beta z}$ does not cause convergence problems if the density of states has bounded support, as in most random matrix ensembles.

In particular, the late time asymptotics of $\langle A(t) A(0) \rangle_\beta$ are still governed by the decomposition Eq.~\eqref{eq:Gxy_approx} of the exact $G(w,z)$. 
Plugging Eq.~\eqref{eq:Gxy_approx} into Eq.~\eqref{eq:finite_temperature_cauchy_transform}, we find the analog of Eq.\eqref{eq:long_time_form_g}:
\begin{equation}
\label{eq:finite_temperature_long_time_g}
  \avg{A(t) A(0)}_\beta \underset{t\to\infty}{\sim} A_{\infty,\beta}^2  + \frac{C_{A,H} }{Z} \avg{e^{(it-\beta)H}} \avg{e^{-iHt}}
\end{equation}
The late time constant $A_{\infty,\beta}^2$ is the same as Eq.\eqref{eq:constant} reweighted by the Boltzmann weight,
\begin{equation}
    A_{\infty,\beta}^2 = \int dx \frac{e^{-\beta x}}{Z} \rho_H(x) \bigg(\text{Re} \Sigma_A(x) + \phi(x) \frac{\text{Im}\Sigma_A(x)}{\pi \rho_H(x)}  \bigg)^2
\end{equation}

Somewhat more interesting is the approach to the constant: for any finite $\beta$ the power law is the same as in infinite temperature case, while in the limit $\beta \to \infty$ the decay is slower since
\begin{equation}
    \frac{1}{Z} \avg{e^{(it-\beta)H}} \to e^{it E_0}
\end{equation}
where $E_0$ is the ground state energy.
Thus, if the finite temperature relaxation follows a power law $1/t^\alpha$, the zero temperature system relaxes as $1/t^{\alpha/2}$.

\section{The Gaussian Rotation Approach}
\label{sec:orthogonal_transformation}
If $A$ and $B$ are both Gaussian, there is an alternative approach that lets us compute any real--time correlator based on an orthogonal transformation. 

When $A$ and $H$ are freely independent, we can easily compute the two--point function
\begin{equation}
  \avg{A(t)A(0)} = \avg{e^{iHt} A e^{-iHt} A}
\end{equation}
using the non--crossing rules between $A$ and $H$, but the problem is of course that we are interested in the case when they are not. 

In section~\ref{sec:exact_analysis} we kept $A$ fixed and expanded $H$ in terms of $A$ and $B$, which do have a non--crossing rule. Here we do the opposite: we hold $H$ fixed, and seek a change of variables that turns $A$ into something freely independent with $H$.
This is easily done in the Gaussian case, but it is not clear
how to construct such a transformation for general ensembles.

As in sec.\ref{sub:gaussian_correlator}, we use traceless matrices with second moments
\begin{equation}
  \{ A^2 \} = \lambda \qquad \{ B^2 \} = 1 - \lambda
\end{equation}
and it is convenient to extract the $ \lambda$ dependence defining the unit variance variables
\begin{equation}
    \tilde{A} = \frac{A}{ \sqrt{\lambda}} \qquad \tilde{B} = \frac{B}{ \sqrt{1-\lambda}}
\end{equation}
to make the algebra in the rest of the section a little cleaner.

We define the variable $C$ through the orthogonal transformation
\begin{equation}
  \begin{pmatrix}
    C \\
    H
  \end{pmatrix}
  =
  \begin{pmatrix}
    \sqrt{1 - \lambda} & -\sqrt{\lambda} \\
    \sqrt{\lambda} & \sqrt{1 - \lambda}
  \end{pmatrix}
  \begin{pmatrix}
    \tilde{A} \\
    \tilde{B}
  \end{pmatrix}
\end{equation}
which makes it Gaussian and independent with $H$. In the $N \to \infty$ limit, independent Gaussian variables become also freely independent\cite{nicaLecturesCombinatoricsFree2006}, so we have found a variable with the requested non--crossing rule with $H$.

The calculation of the correlator at this point is straightforward: we express $A$ as
\begin{equation}
  \tilde{A} = \frac{A}{ \sqrt{\lambda}} = \sqrt{1 - \lambda} C + \sqrt{\lambda} H
\end{equation}
and using the free cumulant expansion we get
\begin{align}
  \frac{\avg{A(t)A(0)}}{\lambda} &= \avg{\tilde{A}(t)\tilde{A}(0)}  \\
  &= \lambda \avg{H^2} + (1-\lambda) \avg{C^2} \abs{\avg{e^{iHt}}}^2 \label{eq:gauss_asymp_form} \\
       &= \lambda + (1-\lambda) \left( \frac{J_1(2t)}{t} \right)^2
\end{align}
which correctly reproduces the result we obtained through the diagrammatic formalism (eq.~\ref{eq:gaussian_time_corr}).

\begin{figure}
  \includegraphics[width=\linewidth]{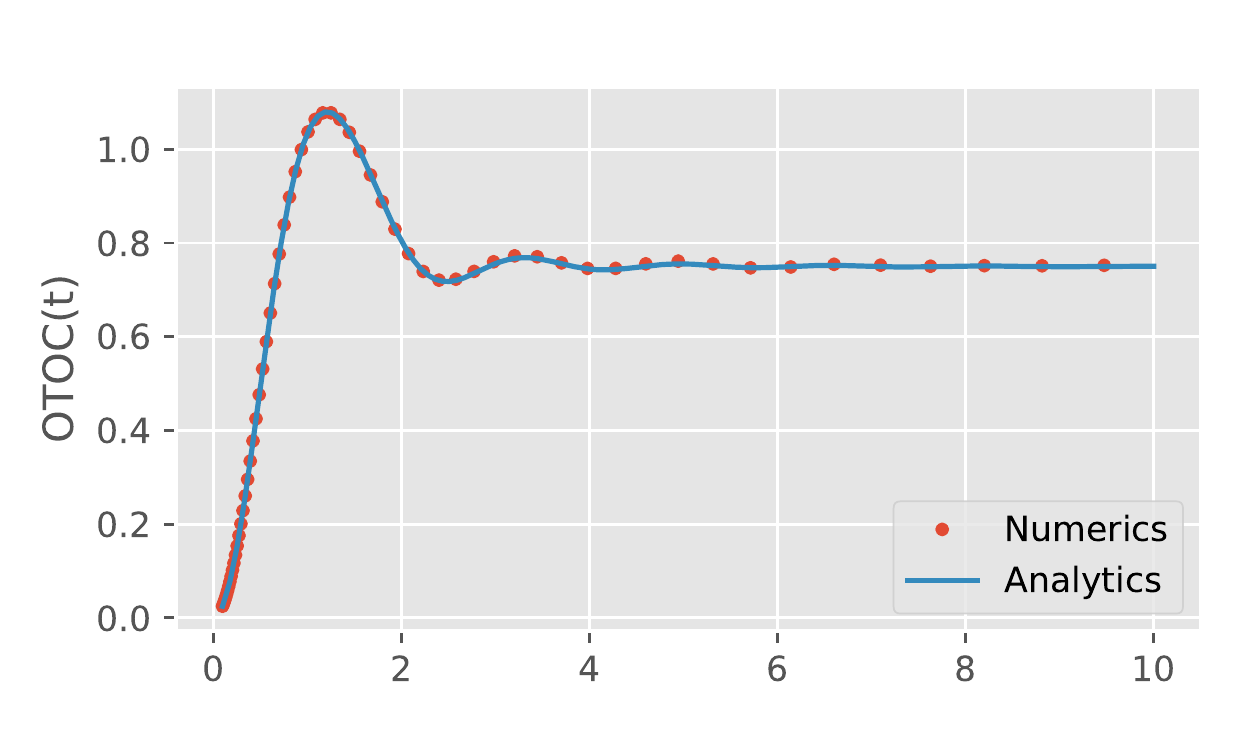}
  \caption{Gaussian OTOC, $N = 200$, disorder average over 10 samples. The two curves in the upper panel are hard to resolve because their difference is less than $10^{-3}$ at all times.}
  \label{fig:otoc_num}
\end{figure}

Using this method we can actually compute any correlator of Gaussian operators easily. For example, the out-of-time-order correlator of $A$ with itself:
\begin{equation}
  \begin{split}
  \text{OTOC}(t) & \coloneqq \frac{1}{2} \avg{\abs{[A(t),A]}^2} \\
  & = \avg{A(t)^2 A^2} - \avg{A(t)AA(t)A}
  \end{split}
\end{equation}
The free cumulant expansion gives in this case, 
\begin{equation}
  \begin{split}
  \frac{\text{OTOC}(t)}{\lambda^2} = 1 - \lambda^2 &+ \left(\frac{J_1(2t)}{t} \right)^2 (1-\lambda)(1-9\lambda) \\
   &- \left(\frac{J_1(2t)}{t}\right)^2 \frac{J_1(4t)}{t} (1-\lambda)^2 \\
   &+ \frac{J_1(2t)}{t} \frac{J_2(2t)}{t^2} 12 \lambda (1-\lambda) \\
   &+  \left(\frac{J_2(2t)}{t}\right)^2 8 \lambda (1-\lambda)
  \end{split}
\end{equation}

This expression is in good agreement with numerics, as we can see from
fig.~\ref{fig:otoc_num}.
The short time behavior is compatible with what is known in literature\cite{vijayFiniteTemperatureScramblingRandom2018}
\begin{equation}
  \frac{\text{OTOC}(t)}{\lambda^2}  = 5(1- \lambda) t^2 + O(t^3)
\end{equation}
but we see a modification of the exponent in the long--time power law: if $A$ is independent of $H$, previous work\cite{vijayFiniteTemperatureScramblingRandom2018} finds a $t^{-4}$ decay to the infinite
time value. The partial conservation of $A$ leads to a slower $1/t^3$ decay:
\begin{equation}
  \begin{split}
  \frac{\text{OTOC}(t)}{\lambda^2} = &1 - \lambda^2 + \frac{(1 - \lambda)^2}{2\pi t^3} \\
  &- (1 - \lambda)(1 - 17 \lambda) \frac{\sin(4t)}{2 \pi t^3} + O\left( \frac{1}{t^4} \right)
  \end{split}
\end{equation}

\section{Numerical Confirmation} 
\label{sec:numerical_confirmation}

\begin{figure*}[t]
    \centering
    \includegraphics[width=\linewidth]{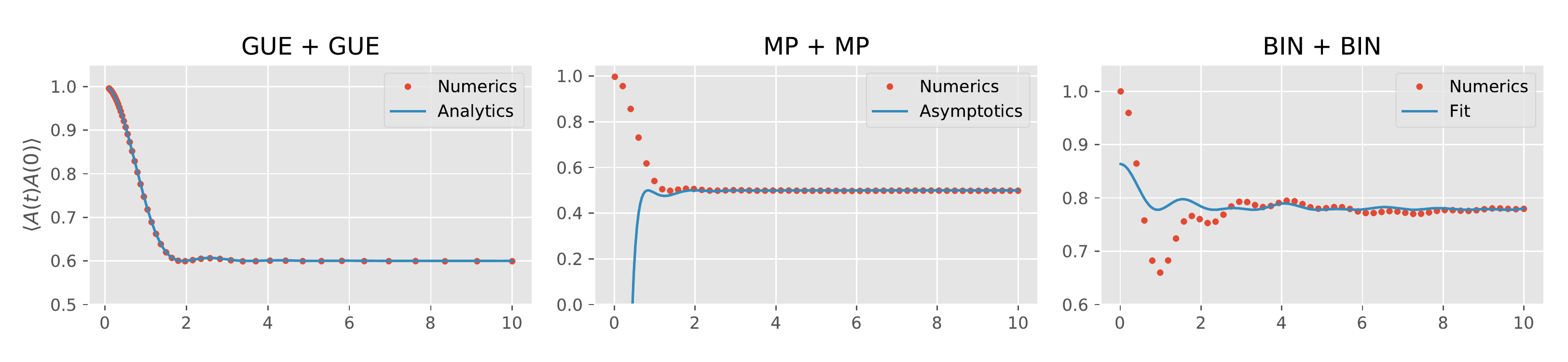}
    \caption{Numerical verification of the prediction of Eq.\eqref{eq:ng_freq_corr} with $N = 500$. See Sec.\ref{sec:numerical_confirmation} for details.}
    \label{fig:comparison_numerics}
\end{figure*}

In this section we check the predictions of eq.~\ref{eq:ng_freq_corr} in a few interesting cases. Table~\ref{tab:numerics_summary} contains a summary of the relevant functions and the resulting long--time constant, while fig.~\ref{fig:comparison_numerics} compares the analytical prediction with the numerical results.

In all three examples, $A$ and $B$ are sampled from the same ensemble simply because the resulting expressions are cleaner, but this is not necessary: our results works just as well in the mixed case.  

\begin{table*}
    \centering
    \caption{Summary of functions and result for a few examples we checked numerically.}
    \label{tab:numerics_summary}
    \setlength{\tabcolsep}{6pt}
    \begin{tabular}{ccCCC} \toprule
    $A$ & $B$ & R_A(z)                    & f_H(z)                                        & \avg{A(t)A(0)}  \\ \midrule
    GUE & GUE & \lambda z                 & \frac{1}{2} \left( z - \sqrt{z^2 - 4} \right) & \lambda  \\ \midrule
    MP  & MP  & \frac{z}{1-z}             & \frac{z+1-\sqrt{z^2-2z-7}}{2(z+2)} & \frac{1}{2} \\ \midrule
    BIN & BIN & \frac{z}{z^2 - \lambda^2} & \frac{z}{\sqrt{(z^2 - (\lambda + \mu)^2)(z^2 - (\lambda - \mu)^2)}} & \max{ \left( \frac{1}{2}, 1 - \frac{ \mu^2}{2 \lambda^2} \right)}\\
    \bottomrule
    \end{tabular}
\end{table*}

\paragraph{GUE}
\label{par:gaussian}
We used the conventions of section~\ref{sub:gaussian_correlator}. 

\paragraph{Marchenko--Pastur (MP)}
\label{par:marchenko_pastur}
To sample an MP matrix $A$ we first sample a standard $GUE$ matrix $M$ with
\begin{equation}
    \avg{M} = 0 \quad \avg{M^2} = 1  
\end{equation}
and then we compute
\begin{equation}
    A = M^2 - \avg{M^2}.
\end{equation}
The same procedure is repeated for the matrix $B$. 

\paragraph{Binary Matrices (BIN)}
\label{par:binary_matrices}
These are matrices of the form
\begin{equation}
    A = \lambda \ U_A^\dagger D U_A \quad B = \mu \ U_B^\dagger D U_B \quad \lambda, \mu > 0
\end{equation}
where $U_i$ are Haar--random $N \times N$ unitaries and $D$ is a diagonal
matrix filled with half $-1$ and half $+1$ values.

While the result eq.~\ref{eq:ng_freq_corr} is still valid for these matrices, the approximation that leads to eq.~\ref{eq:Gxy_approx} breaks down: $ \rho_H(x)$ has $x^{-1/2}$ edges (see the second column in Table~\ref{tab:numerics_summary}). 
This means that the value of the long--time constant eq.\ref{eq:constant} is correct, as confirmed by numerics, but determining the approach requires more work.
    
We instead numerically compute the characteristic function $ \avg{e^{i H t}}$ and fit 
\begin{equation}
    \avg{A(t)A(0)} = \avg{A(\infty)A(0)} + c \abs{ \avg{e^{i H t}} }^2
\end{equation}
to the numerics. Figure \ref{fig:comparison_numerics} shows that this approximation is not as clean as in the other two cases, but after a short time the error settles to $1/N$, which is the best we can hope for. 

This suggests that even though the approximation of eq.~\ref{eq:Gxy_approx} is not valid in this case, the approach to the constant is still determined by the characteristic function of~$H$.

\section{Discussion} 
\label{sec:concluding_thoughts}

There are two directions along which it would be interesting to extend the current work to local, finite dimensional, ergodic quantum systems.

First, if $A$ is a local operator and $B$ is a sum of local operators which itself satisfies ETH, then we expect $H = A+B$ to satisfy ETH and the observable $A$ to be partially conserved. This implies that $\langle A(t) A(0)\rangle \to c/L$ where $L$ is the size of the extended system and $c$ is a constant quantifying how conserved $A$ is. 
This can be computed explicitly using Eq.~\eqref{eq:constant} and the results tested against ETH systems.

More technically challenging is to extend the analysis here to chains of locally interacting random matrices where one might hope to compute the energy diffusion constant explicitly from the dynamical correlators. 
Here the exact resummations available in the random matrix case are complicated by the locality structure of the chain. Some technical steps along this axis have been developed in Ref.~\onlinecite{morampudiManybodySystemsRandom2018}. 

The order of limits is important: $ N \to \infty$ must be taken before $t \to \infty$. 
At finite $N$, we expect corrections of order $1/N$ to the late time value of $\langle A(t) A(0) \rangle_c$, though we have not computed them.
They can be calculated perturbatively by resumming diagrams that tessellate a torus with a hole. 
For $A$ independent of $H$, the $1/N$ corrections can be computed non-perturbatively from dephasing the spectral representation:
\begin{align}
    \langle A(t) A(0) \rangle \to \frac{1}{N} \mathbb{E} \sum_{\alpha} |A_{\alpha \alpha}|^2 = \frac{1}{N} \langle A^2 \rangle
\end{align}
where $\alpha$ runs over the energy eigenbasis.
However, for $A$ part of $H$, this `diagonal ensemble' calculation is not straightforward, as the $\ket{\alpha}$ are correlated with $A$.
Indeed, these correlations must produce both the $O(1/N^0)$ late time value which we have computed and any $O(1/N)$ corrections.

\begin{acknowledgments}
The authors would like to thank A. Chandran and A. Polkovnikov for stimulating discussions. 
C.R.L. acknowledges support from the NSF through grant PHY-1752727.
This work was performed in part at the Aspen Center for Physics, which is supported by National Science Foundation grant PHY-1607611, and at the Galileo Galilei Institute in Florence.
Any opinion, findings, and conclusions or recommendations expressed in this material are those of the authors and do not necessarily reflect the views of the NSF.
\end{acknowledgments}

\appendix

\section{Short time expansion}
\label{sec:short_time_expansion}
In the GUE case, the two--point correlation function can also be computed summing its short time
expansion: we write $G(t)$ as a power series in the Liouvillian hyperoperator~$\mathcal{L}\coloneqq[H,\cdot]$.
\begin{equation}
    \frac{\avg{A(t)A(0)}}{\lambda} = \avg{ \left( e^{i t \mathcal{L}} \tilde{A} \right) \tilde{A}} = \sum_{ n=0}^\infty \frac{(i t)^n}{n !} \avg{ \left( \mathcal{L}^n \tilde{A} \right) \tilde{A}}
\end{equation}
and evaluate explicitly $\avg{ \left( \mathcal{L}^n \tilde{A} \right) \tilde{A}}$. 

These moments exhibit a clear pattern \cite{sloaneSequenceA005568}: for odd powers of $\mathcal{L}$ the expression vanishes, while for even powers we have
\begin{equation}
    \avg{\left(\mathcal{L}^{2n} \tilde{A}\right) \tilde{A}} =
    \begin{cases}
    1 \quad &\text{if} \ n = 0 \\
    C_n C_{n+1}(1- \lambda) \quad \ &\text{if} \ n > 0
    \end{cases}
\end{equation}
where $C_n$ is the n--th Catalan number.
The short time series is then
\begin{equation}
\begin{split}
   \frac{\avg{A(t)A(0)}}{\lambda} &= 1 - (1 - \lambda) \\
   &+ (1 - \lambda) \sum_{n=0}^\infty \frac{(i t)^{2n}}{(2n)!} C_n C_{n+1}
\end{split}
\end{equation}
and it can be summed: 
\begin{equation}
   \frac{\avg{A(t)A(0)}}{\lambda} = \lambda + (1- \lambda) \left(\frac{J_1(2t)}{t}\right)^2
\end{equation}
which agrees with the result presented in the main text.

\bibliography{biblio}

\end{document}